\documentclass[prd,twocolumn,aps,preprintnumbers,amsmath,amssymb,footinbib,superscriptaddress,showpacs]{revtex4}

\usepackage{color}
\usepackage{graphicx}
\usepackage{tabularx}
\usepackage{txfonts}
\usepackage{amssymb}

\begin{document}

\title{Constraints on Dark Matter Annihilation Cross Section in Scenarios of Brane-World and Quintessence}
\author{Wan-Lei Guo}
\email{guowl@itp.ac.cn}\affiliation{Kavli Institute for Theoretical
Physics China, Key Laboratory of Frontiers in Theoretical Physics,
Institute of Theoretical Physics, Chinese Academy of Science,
Beijing 100190, China}

\author{Xin Zhang}
\email{zhangxin@mail.neu.edu.cn}\affiliation{Department of Physics,
College of Sciences, Northeastern University, Shenyang 110004,
China} \affiliation{Kavli Institute for Theoretical Physics China,
Key Laboratory of Frontiers in Theoretical Physics, Institute of
Theoretical Physics, Chinese Academy of Science, Beijing 100190,
China}
\date{\today}

\begin{abstract}

We investigate the dark matter annihilation in the brane-world and
quintessence scenarios, in which the modified cosmological expansion
rate can enhance the thermal relic density of dark matter. According
to the observed dark matter abundance, we constrain the thermally
averaged annihilation cross section $\langle \sigma v \rangle$ in
these two scenarios. In addition, the big bang nucleosynthesis and
the partial-wave unitarity are also used to place bounds on $\langle
\sigma v \rangle$. It is found that both scenarios can lead to a
large annihilation cross section, so they can be used to explain the
recent PAMELA, ATIC and PPB-BETS anomalies.

\end{abstract}


\maketitle

\section{Introduction}

The existence of dark matter is by now well confirmed \cite{DM}. The
recent cosmological observations have helped to establish the
concordance cosmological model where the present Universe consists
of about 73\% dark energy, 23\% dark matter, and  4\% atoms
\cite{WMAP}. However, in the standard model of particle physics,
there is no candidate for dark matter. Understanding the nature of
dark matter is one of the most challenging problems in particle
physics and cosmology. One particularly attractive class of dark
matter candidates is provided by Weakly Interacting Massive
Particles (WIMPs). So far, we still do not know the WIMP's mass $m$
and the thermally averaged annihilation cross section $\langle
\sigma v \rangle$.


In the standard cosmology, the observed dark matter abundance
$\Omega_D h^2 = 0.1131 \pm 0.0034$ \cite{WMAP} requires $\langle
\sigma v \rangle \approx 3 \times 10^{-26} \, {\rm cm^3 \,
sec^{-1}}$, if the $s$-wave annihilation is dominant \cite{KOLB}. In
this case, $\langle \sigma v \rangle$ is insensitive to the
temperature $T$ of the Universe. In fact, the thermal relic density
of dark matter depends not only on their annihilation cross section
$\langle \sigma v \rangle$, but also on the cosmological Hubble
expansion rate during the era of dark matter production and
annihilation. So, any deviation from the standard cosmology, at this
epoch, will lead to some variation of the thermal relic density of
dark matter. Actually, before the big bang nucleosynthesis (BBN),
the Hubble parameter remains unclear. It is likely that some unknown
mechanisms govern the evolution of the Universe in the epoch. For
example, the brane-world scenario \cite{RS} and the quintessence
scenario with a kination phase \cite{QUIN}, may play a significant
role in the pre-BBN era. In these two scenarios, the enhancement of
the Hubble expansion rate will lead to the enhancement of the
thermal relic density of dark matter \cite{Modified}.


For the observed relic density, the modified Hubble parameter
scenarios possess a larger annihilation cross section than $\langle
\sigma v \rangle \approx 3 \times 10^{-26} \, {\rm cm^3 \,
sec^{-1}}$ in the standard cosmology. Therefore, many dark matter
models can derive the larger parameter space when we choose
different Hubble parameter. On the other hand, some models, which
are excluded in the standard cosmology, can work well in these
scenarios. It is very important for us to derive bounds on the
thermally averaged annihilation cross section $\langle \sigma v
\rangle$. In this paper, we try to give a comprehensive constraint
on the dark matter annihilation cross section in the brane-world and
quintessence scenarios. Besides, the BBN and partial-wave unitarity
can also be used to constrain the parameter space of the above two
scenarios. This paper is organized as follows: In Sec. II, we
investigate the constraints on $\langle \sigma v \rangle$ from the
relic density, BBN and unitarity bounds. We also discuss the recent
PAMELA, ATIC and PPB-BETS anomalies. Some discussions and
conclusions are given in Sec. III.


\section{Constraints on dark matter annihilation cross section}

The evolution of dark matter abundance is given by the following
Boltzmann equation \cite{KOLB}:
\begin{eqnarray}
\frac{d Y}{d x} = - \frac{{\bf s}(x)}{x \, H} \langle \sigma v
\rangle (Y^2 -Y_{EQ}^2) \; , \label{bol}
\end{eqnarray}
where $x \equiv m/T$, $Y \equiv n/{\bf s}(x)$ denotes the dark
matter number density, and $Y_{EQ}$ is the equilibrium (EQ) number
density,
\begin{eqnarray}
Y_{EQ} & = & \frac{45}{4 \pi^4} \frac{g_i}{g_*} x^2 K_2(x) \;,
\end{eqnarray}
with $K_2(x)$ the modified Bessel functions, and $g_*$ the total
number of effective relativistic degrees of freedom. For the
internal degrees of freedom of dark matter particle, we take $g_i
=1$. The entropy density ${\bf s}(x)$ is given by
\begin{eqnarray}
{\bf s}(x)  =  \frac{2 \pi^2 g_*}{45} \frac{m^3}{x^3} \;.
\end{eqnarray}
In the standard cosmology (SC), the Hubble expansion rate $H$ is
written as
\begin{eqnarray}
H_{SC}  =  \sqrt{\frac{4 \pi^3 g_*}{45}} \frac{T^2}{M_{Pl}}
\label{h} \;,
\end{eqnarray}
where $M_{Pl}=1.22 \times 10^{19}$ GeV is the Planck (Pl) mass.
Using the result of Eq.~(\ref{bol}), $Y_0$, one can obtain the dark
matter relic density $\Omega_D h^2$ \cite{APP}:
\begin{eqnarray}
\Omega_D h^2 =2.74 \times 10^8 \frac{m}{\rm GeV} Y_0 \;.
\end{eqnarray}

The thermal average of annihilation cross section times the
``relative velocity", $\langle \sigma v \rangle$, is a key quantity
in the determination of the cosmic relic abundances of dark matter.
In the standard cosmology, one can use approximate formulas to
calculate the dark matter relic density. Then, $\langle \sigma v
\rangle \approx 3 \times 10^{-26} \, {\rm cm^3 \, sec^{-1}}$ can be
obtained for the $s$-wave annihilation  \cite{KOLB, Guo}. If the
Hubble expansion rate $H$ deviates from that of the standard
cosmology, one should numerically resolve the Boltzmann equation in
Eq. (\ref{bol}). In this case, we have three parameters: $m$,
$\langle \sigma v \rangle$ and $H$. For the WIMP, its mass $m$
should be roughly between 10 GeV and a few TeV for annihilation
cross section of approximately weak strength. So, we take $10 \,
{\rm GeV} \leq m \leq 10$ TeV. In this paper, we only focus on the
$s$-wave annihilation of dark matter. Thus, $\langle \sigma v
\rangle$ is assumed to be a constant. In terms of the partial-wave
unitarity, one can derive the unitarity bound on $\langle \sigma v
\rangle$ \cite{Unitarity}:
\begin{eqnarray}
\langle \sigma v \rangle \lesssim \frac{3 \times 10^{-22} \, {\rm
cm^3 \, sec^{-1}}}{(m/ {\rm TeV})^2} \label{unitarity}\;.
\end{eqnarray}
We shall show that this limit will exclude part of the parameter
space. Using the observed dark matter abundance $\Omega_D h^2 =
0.1131 \pm 0.0034$ \cite{WMAP}, we discuss the constraints on the
above three unknown parameters in the following subsections.

\subsection{The brane-world scenario}

In this subsection, we consider the dark matter annihilation in a
brane-world scenario, in which the standard model particles are
assumed to be confined on a 3-brane. In the early Universe,
extra-dimension effects might play an important role. It is of
interest to study how the physics of extra dimensions may affect the
dark matter annihilation.

Focusing on the case with one extra dimension compactified on a
circle, the effective four dimensional Friedmann equation is
\cite{RS}
\begin{eqnarray}
H^2 = \frac{8 \pi}{3 M_{Pl}^2} \rho \left( 1 + \frac{\rho}{\rho_c}
\right) \label{hbc} \;,
\end{eqnarray}
where $\rho = \pi^2 g_* T^4 / 30 $ is the usual energy density of
radiation, and $\rho_c=2\Lambda$ is the critical density, with
$\Lambda$ the brane tension. The critical density $\rho_c$ can be
expressed as
\begin{eqnarray}
\rho_c = \frac{96 \pi M_5^6}{M_{Pl}^2} \;,
\end{eqnarray}
where $M_5$ denotes the true gravity scale of the five dimensional
theory. Here, we have ignored the dark-radiation term in Eq.
(\ref{hbc}), considering the BBN constraint. On the other hand, the
BBN can also constrain the critical density $\rho_c$,
\begin{eqnarray}
\left (\rho/\rho_c \right)_{T = 1 {\rm  MeV}} \leq 1   \;.
\end{eqnarray}
Then, we straightforwardly derive the BBN bound,
\begin{eqnarray}
M_5 \geq 1.1 \times 10^4 \, {\rm GeV} \label{BBNB} \;.
\end{eqnarray}
It should be mentioned that the precise measurements of the
gravitational law in submillimeter scale give a more strict limit
$M_5 > 1.1 \times 10^8$ GeV \cite{Langlois}. However, this
constraint is model dependent \cite{MW}. So we only consider the BBN
bound $M_5 \geq 1.1 \times 10^4$ GeV in the following analysis.

\begin{figure}[t]\begin{center}
\includegraphics[scale=0.5]{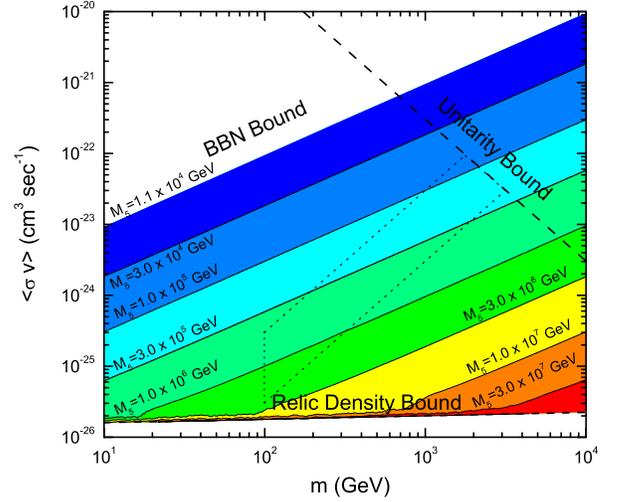}
\end{center}
\caption{The allowed region of $\langle \sigma v \rangle$ and $m$
for different $M_5$ in the brane-world cosmology from the observed
relic density $\Omega_D h^2 = 0.1131 \pm 0.0034$. Here, we also
consider the unitarity bound in Eq. (\ref{unitarity}) and the BBN
bound in Eq. (\ref{BBNB}). The region with dotted boundary denotes
the allowed range of $\langle \sigma v \rangle$ in Eq.
(\ref{pamela}) from the PAMELA experiment. } \label{BC}\end{figure}

Using the modified Friedmann equation Eq. (\ref{hbc}) and the
observed relic density $\Omega_D h^2 = 0.1131 \pm 0.0034$, we
numerically solve the Boltzmann equation. Our numerical results are
shown in Fig.~\ref{BC}. Here, we have chosen $g_* = 106.75$ for
illustration. For most of the parameter range,  one can find
\begin{eqnarray}
\Omega_D h^2 \propto \frac{m}{\langle \sigma v \rangle} M_5^{-3/2}
\;.
\end{eqnarray}
This feature can be approximately derived from the Boltzmann
equation when we omit the equilibrium number density $Y_{EQ}$ of Eq.
(\ref{bol}) and require $\rho/\rho_c \gg 1$ at the usual freeze-out
time of $x_f \approx 20$. In Fig.~\ref{BC}, the relic density bound
describes the standard cosmology case $H = H_{SC}$. In this case,
the relic density bound $\langle \sigma v \rangle \gtrsim 2 \times
10^{-26} \, {\rm cm^3 \, sec^{-1}}$, indicates that the predicted
relic density is not bigger than the observed value. As shown in
Fig.~\ref{BC}, the BBN, unitarity and relic density bounds,
together, strictly constrain the thermally averaged annihilation
cross section,
\begin{eqnarray}
2 \times 10^{-26}  \lesssim \langle \sigma v \rangle \lesssim 3
\times 10^{-22} \, {\min \left[\frac{3 m}{\rm TeV}, \frac{\rm
TeV^2}{m^2}\right]} \;, \label{Bound1}
\end{eqnarray}
where $\langle \sigma v \rangle$ is in units of ${\rm cm^3 \,
sec^{-1}}$.

\subsection{The quintessence scenario}

In this subsection, we consider another possible scenario in which
the dark energy is not negligible in the very early Universe. It is
well-known that in the present Universe, the dark energy has begun
to play a dominant role. Due to the repulsive gravity of dark
energy, the Universe is currently undergoing an accelerated
expansion. However, we still do not know whether the dark energy is
the cosmological constant or some dynamical scalar field. If the
dark energy is some scalar field, for example, the quintessence
field, it is possible that the dark energy can dominate the early
Universe. Of course, in the early Universe, the kinetic term
dominates the energy density of the scalar field.

In the quintessence scenario, the Friedmann equation can be written
as \cite{QUIN}
\begin{eqnarray}
H^2 = \frac{8 \pi}{3 M_{Pl}^2} \rho \left( 1 +
\frac{\rho_\Phi}{\rho} \right) \;,
\end{eqnarray}
where $\rho_\Phi$ is the quintessence dark energy density. The
kinetic term domination leads to
\begin{eqnarray}
\rho_\Phi = \eta \frac{\pi^2}{15} \frac{g_*^2}{10.75^2}
\frac{T^6}{{\rm MeV^2}} \;.
\end{eqnarray}
The parameter $\eta \equiv \left (\rho_\Phi/\rho_\gamma \right)_{T =
1 {\rm MeV}}$ is defined by the ratio of quintessence-to-photon
energy densities at the temperature  $T = 1$ MeV. Considering the
BBN constraint, we require
\begin{eqnarray}
\left (\rho_\Phi/\rho \right)_{T = 1 {\rm  MeV}} \leq 1 \;.
\end{eqnarray}
Then, one can derive the BBN bound for the quintessence scenario,
\begin{eqnarray}
\eta \leq 5.4 \label{BBNQ} \;.
\end{eqnarray}

\begin{figure}[t]\begin{center}
\includegraphics[scale=0.5]{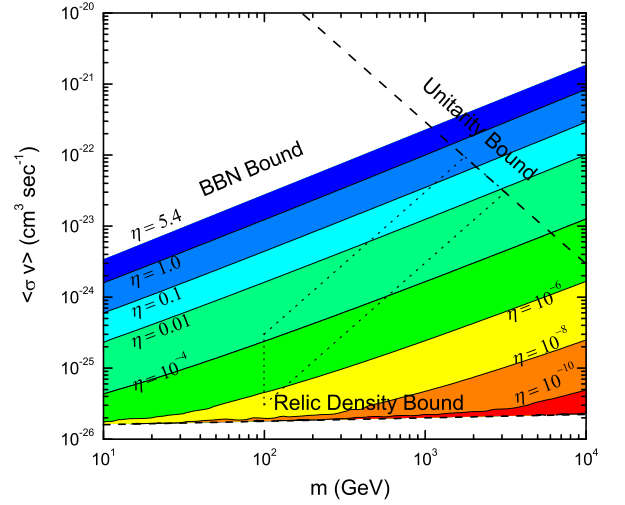}
\end{center}
\caption{The allowed region of $\langle \sigma v \rangle$ and $m$
for different $\eta$ in the quintessence scenario from the observed
relic density $\Omega_D h^2 = 0.1131 \pm 0.0034$. Here, we also
consider the unitarity bound in Eq. (\ref{unitarity}) and the BBN
bound in Eq. (\ref{BBNQ}). The region with dotted boundary denotes
the allowed range of $\langle \sigma v \rangle$ in Eq.
(\ref{pamela}) from the PAMELA experiment. } \label{DE}\end{figure}

If $\rho_\Phi/\rho \ll 1$  at the usual freeze-out time $x_f \approx
20$ ($m^2 \eta \ll 2 \times 10^{-4}$, $m$ is in unit of GeV), one
can totally omit the dark energy contribution. In this case, one can
derive the standard cosmology bound $\langle \sigma v \rangle
\approx 3 \times 10^{-26} \, {\rm cm^3 \, sec^{-1}}$. However, if
the quintessence kinetic energy density dominates the early
Universe, $\langle \sigma v \rangle$ can be enhanced significantly.
Our numerical results are shown in Fig.~\ref{DE}. For most of the
parameter space, there is a relation between the dark matter mass
$m$ and annihilation cross section $\langle \sigma v \rangle $:
\begin{eqnarray}
\Omega_D h^2 \propto \frac{m}{\langle \sigma v \rangle }
\frac{\sqrt{\eta}}{\log (m \times C)}\;,
\end{eqnarray}
where $C \sim  \sqrt{\eta} \times 10^6$ and $m$ is in units of TeV.
This feature can also be approximately deduced from the Boltzmann
equation if one ignores the equilibrium number density $Y_{EQ}$ and
requires $m^2 \eta \gg 2 \times 10^{-4}$. One can see that the BBN,
unitarity and relic density bounds can effectively constrain the
thermally averaged annihilation cross section $\langle \sigma v
\rangle $. For $10 \, {\rm GeV} \leq m \leq 10$ TeV, we obtain
\begin{eqnarray}
2 \times 10^{-26}  \lesssim \langle \sigma v \rangle \lesssim
3\times 10^{-22} \min \left[ \frac{4 m}{6 + \log (m)},
\frac{1}{m^2}\right] \;, \label{Bound2}
\end{eqnarray}
where $m$ is in units of TeV  and $\langle \sigma v \rangle$ is in
units of ${\rm cm^3 \, sec^{-1}}$.

\subsection{The PAMELA, ATIC and PPB-BETS Anomalies}

Recently, the indirect dark matter detection experiment PAMELA
\cite{PAMELA} reported an excess in the positron fraction from 10 to
100 GeV, but showed no excess for the antiproton data.  Dark matter
annihilation can account for the PAMELA experiment if \cite{BF}
\begin{eqnarray}
BF \times \langle \sigma v \rangle \sim 3 \times 10^{-23}
\left(\frac{m}{\rm TeV}\right)^2 \, {\rm cm^3 \, sec^{-1}} \;,
\end{eqnarray}
where $BF$ is the boost factor. Current analysis on the clumpiness
of dark matter structures indicates that the biggest probable boost
factor should be less than $10-20$ \cite{Structure}. Therefore, we
have
\begin{eqnarray}
\langle \sigma v \rangle \sim 3 \times 10^{-24} - 3 \times 10^{-23}
\, \left(\frac{m}{\rm TeV}\right)^2 \, {\rm cm^3 \, sec^{-1}}
\;.\label{pamela}
\end{eqnarray}
In the standard cosmology, such a large $\langle \sigma v \rangle$
will give a smaller dark matter abundance than the observed one. So
far, many authors have provided several very interesting mechanisms
to try to resolve the relic density puzzle, including the Sommerfeld
enhancement \cite{Sommerfeld} and the Breit-Wigner enhancement
\cite{BW} (for the relevant discussions, see  Ref. \cite{BW1}). In
these mechanisms, $\langle \sigma v \rangle$ will increase with the
expansion of the Universe. If $\langle \sigma v \rangle$ is a
constant, one must resort to the non-thermal production of dark
matter \cite{Zupan}. In the cuspy halo profile case, the
annihilation cross section in Eq. (\ref{pamela}) will produce
abundant gamma rays from galactic center, which conflicts with the
HESS results \cite{Gamma}. In fact, current observations do not give
the direct evidence that any nearby galaxy has a cusped dark matter
halo profile. In addition, the long-lived intermediate state can
relax the gamma ray constraint \cite{Longlive}.

In the brane-world scenario, the authors of Ref. \cite{Zant}
approximately estimate that the large expansion rate $H$ may admit
$\langle \sigma v \rangle \sim 10^{-6} \,{\rm GeV}^{-2}$, which can
accommodate the PAMELA results and the correct relic density. In
this paper, we have numerically calculated the Boltzmann equation
and constrained $m$, $\langle \sigma v \rangle$ and $M_5$ (or
$\eta$) from the PAMELA results in both brane-world and quintessence
scenarios. As shown in Figs.~\ref{BC} and \ref{DE} (the region with
dotted boundary), the required $\langle \sigma v \rangle$ in Eq.
(\ref{pamela})  can produce the observed dark matter abundance. It
is worthwhile to stress that the unitarity bound indicates $m
\lesssim 3$ TeV for the PAMELA result. In addition to the PAMELA
experiment, the ATIC \cite{ATIC} and PPB-BETS \cite{PPB} balloon
experiments have also seen the excess in the $e^+ + e^-$ energy
spectrum between 300 and 800 GeV. In order to explain the ATIC and
PPB-BETS anomalies, the dark matter mass $m$ should be order of TeV.
It is clear that the brane-world and quintessence scenarios can
simultaneously explain the dark matter relic density and the PAMELA,
ATIC and PPB-BETS anomalies.


\section{Discussion and Conclusion}

We have constrained the dark matter thermally averaged annihilation
cross section $\langle \sigma v \rangle$ in the brane-world and
quintessence scenarios. In these two scenarios, the Hubble expansion
rate can be written as
\begin{eqnarray}
H = H_{SC} \sqrt{1 + Z(T)} \;,
\end{eqnarray}
with
\begin{eqnarray}
Z(T) & \equiv & \frac{\rho}{\rho_c}  =  \frac{\pi g_* }{2880 }
\frac{M_{Pl}^2}{M_5^6} \frac{T^4}{{\rm GeV}^4}  \;; \; ({\rm Brane
\; world})\\ Z(T) & \equiv & \frac{\rho_\Phi}{\rho}  =  \frac{2 \eta
g_* 10^6}{10.75^2} \frac{T^2}{{\rm GeV}^2} \;. \; ({\rm
Quintessence})
\end{eqnarray}
It should be emphasized that our previous analysis in Sec. II can be
applied to other models in which $Z(T)$ is proportional to $T^4$ or
$T^2$.

In conclusion, in the scenarios of brane-world and quintessence, the
modified Hubble expansion rate $H$ can enhance the thermal relic
density of dark matter. Furthermore, the constraint on the thermally
averaged annihilation cross section $\langle \sigma v \rangle
\approx 3 \times 10^{-26} \, {\rm cm^3 \, sec^{-1}}$ can be relaxed
for the $s$-wave annihilation. Considering the relic density, BBN
and unitarity bounds, we have derived the general constraints on
$\langle \sigma v \rangle$ in Eqs. (\ref{Bound1}) and (\ref{Bound2})
for the brane-world and quintessence scenarios, respectively. As
shown in Figs. \ref{BC} and \ref{DE},  the two scenarios discussed
in this paper can simultaneously explain the observed dark matter
abundance $\Omega_D h^2 = 0.1131 \pm 0.0034$ and  the PAMELA, ATIC,
and PPB-BETS anomalies. It should be mentioned that the
observational data from the dark matter may be used to constrain
cosmological models. Once the dark matter mass $m$ and annihilation
cross section $\langle \sigma v \rangle $ are fixed by the dark
matter search experiments, we can probe the very early stage of the
Universe from our numerical results in Figs. \ref{BC} and \ref{DE}.

\acknowledgments

This work was supported by the National Natural Science Foundation
of China (NSFC) under Grants No. 10847163 and No. 10705041.


\begin{thebibliography}{99}



\bibitem{DM} For reviews, see, e.g., G. Jungman, M. Kamionkowski and K.
Griest, Phys. Rept. {\bf 267}, 195 (1996); G. Bertone, D. Hooper and
J. Silk, Phys. Rept. {\bf 405}, 279 (2005).

\bibitem{WMAP} E. Komatsu {\it et al.}, Astrophys. J. Suppl. Ser. {\bf 180}, 330 (2009).

\bibitem{KOLB} E. W. Kolb and M. S. Turner, {\it The Early Universe}
(Addison-Wesley, Reading, MA, 1990).

\bibitem{RS} L. Randall and R. Sundrum,  Phys. Rev. Lett. {\bf 83}, 4690
(1999).

\bibitem{QUIN} P. Salati, Phys. Lett. B {\bf 571}, 121 (2003).


\bibitem{Modified} F. Rosati, Phys. Lett. B {\bf 570}, 5 (2003);
N. Okada and O. Seto, Phys. Rev. D {\bf 70}, 083531 (2004); Phys.
Rev. D {\bf 71}, 023517 (2005); T. Nihei, N. Okada and O. Seto,
Phys. Rev. D {\bf 71}, 063535 (2005);  C. Pallis, JCAP {\bf 0510},
015 (2005); M. Schelke, R. Catena, N. Fornengo, A. Masiero and M.
Pietroni, Phys. Rev. D {\bf 74}, 083505 (2006);  E. Abou El Dahab
and S. Khalil, JHEP {\bf 0609}, 042 (2006); M. Drees, H. Iminniyaz
and M. Kakizaki, Phys. Rev. D {\bf 76}, 103524 (2007); C. Bambi and
F. R. Urban, Phys. Rev. Lett. {\bf 99}, 191302 (2007); A. Arbey and
F. Mahmoudi, Phys. Lett. B {\bf 669}, 46 (2008).


\bibitem{APP}  P. Gondolo and G. Gelmini, Nucl. Phys. {\bf B360}, 145
(1991).

\bibitem{Guo} W. L. Guo, L. M. Wang,
Y. L. Wu and C. Zhuang, Phys. Rev. D {\bf 78}, 035015 (2008); W. L.
Guo, L. M. Wang, Y. L. Wu, Y. F. Zhou and C. Zhuang, Phys. Rev. D
{\bf 79}, 055015 (2009).


\bibitem{Unitarity}  K. Griest and M. Kamionkowski, Phys. Rev. Lett. {\bf 64}, 615 (1990);
L. Hui, Phys. Rev. Lett. {\bf 86}, 3467 (2001).

\bibitem{Langlois} D. Langlois, Prog. Theor. Phys. Suppl. {\bf 148},
181 (2002), and references therein.


\bibitem{MW} K. I. Maeda and D. Wands, Phys. Rev. D {\bf 62}, 124009 (2000).


\bibitem{PAMELA} O. Adriani {\it et al.},
Nature {\bf 458}, 607 (2009); Phys. Rev. Lett. {\bf 102}, 051101
(2009).



\bibitem{BF} M. Cirelli, M. Kadastik, M. Raidal and A. Strumia,
Nucl. Phys.  {\bf B813}, 1 (2009); P. F. Yin, Q. Yuan, J. Liu, J.
Zhang, X. J. Bi, S. H. Zhu and X. M. Zhang, Phys. Rev. D {\bf 79},
023512 (2009).


\bibitem{Structure} J. Lavalle, Q. Yuan, D. Maurin and X. J. Bi,
Astron. Astrophys. {\bf 479}, 427 (2008).

\bibitem{Sommerfeld} J. Hisano, S. Matsumoto, M.
Nagai, O. Saito and M. Senami, Phys. Lett. B {\bf 646}, 34 (2007);
M. Cirelli, A. Strumia, M. Tamburini, Nucl. Phys. {\bf B787}, 152
(2007);  N. Arkani-Hamed, D. P. Finkbeiner, T. R. Slatyer and N.
Weiner, Phys. Rev. D {\bf 79}, 015014 (2009); M. Lattanzi and J.
Silk, Phys. Rev. D {\bf 79}, 083523 (2009).

\bibitem{BW} M. Ibe, H. Murayama and T. Yanagida,
Phys. Rev. D {\bf 79}, 095009 (2009); W. L. Guo and Y. L. Wu, Phys.
Rev. D {\bf 79}, 055012 (2009).

\bibitem{BW1} M. Pospelov and A. Ritz, Phys. Lett. B {\bf 671}, 391 (2009); D. Feldman, Z. Liu and P.
Nath, Phys. Rev. D {\bf 79}, 063509 (2009); J. March-Russell and S.
M. West, Phys. Lett. B {\bf 676}, 133 (2009);  X. J. Bi, X. G. He
and Q. Yuan, arXiv:0903.0122.

\bibitem{Zupan} W. B. Lin, D. H. Huang, X. Zhang and R. H. Brandenberger,
Phys. Rev. Lett. {\bf 86}, 954 (2001); M. Fairbairn and J. Zupan,
arXiv:0810.4147.

\bibitem{Gamma} M. Cirelli and P. Panci, arXiv:0904.3830;
P. Meade, M. Papucci, A. strumia and T. Volansky, arXiv:0905.0480;
and references therein.

\bibitem{Longlive} I. Z. Rothstein, T. Schwetz and J. Zupan, arXiv:
0903.3116.

\bibitem{Zant} A. A. El Zant, S. Khalil and H. Okada,
arXiv:0903.5083.

\bibitem{ATIC} J. Chang {\it et al.}, Nature {\bf 456}, 362 (2008).

\bibitem{PPB} S. Torii {\it et al.}, arXiv:0809.0760.

\end{thebibliography}
\end{document}